\documentstyle[12pt]{article}
\newcommand{\be}{\begin{equation}}
\newcommand{\ee}{\end{equation}}
\newcommand{\nn}{\nonumber}
\begin{document}
\hfill IFT-94/14
\begin{center}
{\Huge New probes of anomalous $WW\gamma$
couplings at future $e^{+}e^{-}$ Linacs
}\footnote{Talk presented at the XVII Warsaw Symposium on
Elementary Particle Physics, Kazimierz, May 23-27, 1994} \\

\vspace{1cm}
K.J. Abraham\footnote{Supported by the grant 2 0417 91 01 from Polish
Committee for Scientific Research}, J.
Kalinowski\footnote{Supported in part by the grant 2 P302 095 05
from Polish Committee for Scientific Research}, \& P. \'Sciepko \\
Insitute of Theoretical Physics \\
ul. Ho\.{z}a 69  \\
00 681 Warsaw  \\
POLAND
\end{center}
\vspace{1cm}
\begin{abstract} We investigate the sensitivity of single photon plus
missing energy
cross-sections at future $e^{+}e^{-}$ linacs to non-standard $WW\gamma$
couplings. We show that even with conservative estimates of systematic errors
there is still considerable sensitivity to anomalous couplings.
Analytic expressions for helicity amplitudes are presented.
\end{abstract}
\newpage
One of the primary physics goals of the next linear collider will be
a detailed investigation of the non-abelian sector of the standard model,
in particular to improve LEP bounds on non-Yang-Mills like triple gauge boson
vertices. Most studies uptill now have focussed on the reaction
$e^{+}e^{-} \rightarrow W^{+}W^{-}$ \cite{conf},
which despite being a sensitive probe of
non-standard $WW\gamma$ and $WWZ$ couplings sufers from the drawback that there
is no obvious way to disentangle the effects of $WW\gamma$ and $WWZ$
form-factors. Hence it would be desirable to investigate other channels where
such $\gamma-Z$ interference effects are not present.

In this letter we suggest that
the process $e^{+}e^{-}\rightarrow \overline{\nu}\nu\gamma$
may be used to study the precise structure of
the $WW\gamma$ vertex without complications from intereference effects
described above. Such final states have been studied at PETRA \cite{Karel}
and at LEP \cite{frits} as
a means of determining the number of light neutrini. At these energies though,
not much sensitivity to anomalous $WW\gamma$ form factors should be expected,
as s-channel processes mediated by virtual Z exchange also play an
important role. However, at Next Linear Collider (NLC) energies
($\sqrt{s} = 500 GeV$), s channel contributions
become less important and the bulk of the cross-section comes from t channel
W exchange. As we will show, it is possible to choose cuts which enhance the
sensitivity of the observed cross-sections and differential distributions to
the $WW\gamma$ form-factor.

How does one parameterise possible deviations from the standard model non
abelian vertices ?
As was shown by Hagiwara et.al \cite{HPZH}, there are seven possible Lorentz
and $U(1)_{em}$ invariant triple gauge boson form-factors. If we require
C and P invariance then only two, traditionally denoted by $\kappa$ and
$\lambda$, remain. Neglecting the form factors violating C and P is reasonable
for our purposes as in the absence of beam polarisation
there is no way to detect CP violating asymmetries if the only particle
detected is a photon of unknown polarisation.
We assume that physics at a scale much higher than those
directly accesible by experiment is responsible for these form factors. With
this assumption, we may set $\kappa$ and $\lambda$ to be constant.  In the
standard model we have $\kappa =1$ and $\lambda = 0 $. Deviations from the
standard model are then parameterised by $\delta \kappa = \kappa -1 $ and
$\lambda$.  The modified Feynman rules for the $WW\gamma$ vertex may be
obtained from
\cite{yehudai}, all other Feynman rules are standard ones.

Due to the complexity of the Feynman rules for the non-standard couplings
it turns out to be convenient to calculate the matrix-element using the
helicity amplitude formalism \cite{CALKUL,gk}. The results are presented in
the Appendix. Note that the helicity amplitudes for non-standard couplings
have not been presented elsewhere. The helicity amplitudes for the
standard model agree with those in Appendix A of \cite{frits}.
Note however, that some of the analytical expressions in \cite{frits} are
incorrect, in particular not all the terms in Eq. 3 have the same dimension.
As a check we have independently evaluated the helicity amplitudes using
the formalisms of \cite{Hagizep} and \cite{gk}
and find excellent numerical agreement
for various values of $\kappa$ and $\lambda$.
As a further check we have verified the results for the differential
distributions at the Z resonance presented in \cite{frits}.

We are now in a position to calculate
$\sigma(e^{+}e^{-} \rightarrow \overline{\nu}\nu\gamma)$ at NLC energies with
non-standard $WW\gamma$ form factors. Since we assume the electron to be
massless we need to impose a minumum angle cut on the direction of the
outgoing photon to avoid colinear singularities as well as a minimum
energy cut to avoid IR problems. Setting $\theta_{min} = 20^o$ and
$E_{min} = 10 GeV$ we find for the Standard Model
a cross-section of $\sim 1.6$ pb, which at
projected NLC luminosities ($\sim 10$ fb$^{-1}$) represents a sizable
number of events. However, with these cuts alone the sensitivity to
non-standard couplings is rather small. This is hardly surprising; the bulk
of the cross-section comes from initial state soft photon bremstrahlung and
is thus only weakly dependent on the non-abelian couplings. Further cuts
are required in order to enhance the relative importance of the $WW\gamma$
couplings.
As the anomalous form factors are associated with higher dimensional
operators containing derivative interactions, it is clear that only the more
energetic photons will be sensitive to the additional derivatives. Therefore
we require that the minumum energy of the photon be 80 GeV. In order to further
improve the situation it is necessary to reduce the background from the Z
exchange graphs. This can easily be achieved by means of a simple kinematical
trick.  In the limit that the Z width is negligible, the photon is essentially
monochromatic, with an energy close to half the centre of mass energy. However,
this corresponds to almost the edge of phase space, where the cross-section
for the t channel graphs is close to zero. Hence we require that the energy
of the photon be less than 180 GeV, which not only removes the background from
Z exchange graphs but also does not reduce the signal from the WW
exchange graphs too much.

With these cuts on the photon energy, the cross-section for the
standard model is 0.21 pb, which still leads to an appreciable number of
events at NLC luminosities. Cross-sections for non-standard values of
$\delta\kappa$ and $\lambda$ with the cuts mentioned above are presented in
Figs. 2 and 3. (We have deliberately varied $\delta \kappa$ and $\lambda$
individually
and not simultaneously, in order to keep the analysis simple.) As is clear
from the curves, there is considerable sensitivity to deviations from the
standard model. Precise discovery limits though, are dependent on what
assumptions are made concerning systematic uncertainties, since
statistical errors are probably quite small given the large numbers of events
$ {\cal O}( 2000)$. Assuming there are no experimental systematic errors, the
main source of theoretical systematic errors lies in unknown higher order
corrections. Note that the higher order corrections discussed in \cite{frits}
are those which are dominant on the Z pole, and are therefore not adequate
for our purposes. It is reasonable to assume that the bulk of the
corrections are real and virtual QED corrections which integrated over the
bulk of phase space are probably quite small. However we are restricting
ourselves to a limited region of the total phase space where
radiative corrections may be sizable even though the total
radiative corrections themselves
are small.
Therefore we make the conservative assumption that the
systematic uncertainties
due to radiative corrections is ${\cal O}( 20 \%) $.

With this assumption it is possible to put the folowing discovery bounds
$ -.6 < \lambda < .6$ and $ -0.6 < \delta \kappa < 2.2$ using the cross-section
with the cuts mentioned above as only sensitive variable. It is interesting to
note that this compares favourably with the bounds obtained by McKellar and He
\cite{He} on the basis of the recent CLEO measurement of
$b \rightarrow s \gamma$ \cite{CLEO}.  Further refinement
is possible if one considers differential distributions. This is illustrated
in Fig. 3 where we have plotted the differential distribution with respect
to photon energy for the standard model and for two values of $\delta
\kappa$. Although the
cross-sections are almost the same the differential
distributions are different. Similar effect is observed for $\lambda$.
It would not be wise though, to derive further discovery bounds on the
basis of differential distributions without a detailed consideration of
detector acceptances and higher order radiative corrections.

To conclude, we have demonstrated that
$\sigma(e^{+}e^{-}\rightarrow \overline{\nu}\nu\gamma)$ at projected NLC
energies and luminosities is sensitive to anomalous $WW\gamma$ couplings.
Making conservative estimates for systematic errors involved it is
possible to constrain $\lambda$ and $\kappa$ to lie within the regions
$ -.6 < \lambda < .6$ and $ -0.6 < \delta \kappa < 2.2$.
These bounds can probably be
improved through knowledge of currently unknown radiative corrections.
We have also for the first time presented all the relevant helicity
amplitudes in a simple and usable form.

Upon completion of this work we became aware of a paper by Couture and
Godfrey OCIP/C 94-4, UQAM-PHE-94-04, addressing similar issues but containing
only numerical and no analytical results. The discovery limits
they derive are much more stringent than ours due to more optmistic assumptions
about the size of theoretical systematic errors.

\bigskip
\noindent{\em Acknowledgments} \\
KJA is indebted to D. Zeppenfeld, H. Murayama
and W.L. van Neerven for valuable clarifications.

\vspace*{6mm}
{\Large \bf Appendix}

\vspace{4mm}
We consider the process $e^+e^-\rightarrow \nu\bar{\nu}\gamma$
with the following kinematics (all momenta are physical, $i.e.$ with
positive energy)
\begin{equation}
e^+(p_1)+e^-(p_2)\rightarrow \nu(p_3)+\bar{\nu}(p_4)+\gamma(p_5)
\nonumber
\end{equation}
The results for the helicity apmlitudes can be conveniently
expressed in terms of inner spinor products defined as follows
\begin{equation}
   <ij>\equiv <p_i-|p_j+>=\sqrt{p_i^+p_j^-}\exp[-i(\omega_i-\omega_j)/2]
 -\sqrt{p_i^-p_j^+}\exp[i(\omega_i-\omega_j)/2],
\end{equation}
where $|p\pm>$ denotes the fermion with momentum $p$ and helicity $\pm$,
$p_i^{\pm}=p_i^0\pm p_i^3$ and $\exp(i\omega_i)=(p_i^1+ip_i^2)/p_i^T$.
The inner spinor product has the following properties \cite{gk}
\begin{eqnarray}
 &<p_i-|\gamma^{\mu}|p_j->=<p_j+|\gamma^{\mu}|p_i+> \nn \\
 &<p_i+|\gamma^{\mu}|p_j+><p_k+|\gamma_{\mu}|p_l+>=2<p_k+|p_i-><p_j-|p_l+>
 &\nn \\
 &<p_i+|p_j->=<p_j+|p_i->^* \nonumber \\
 &|<p_i-|p_j+>|^2=2\, p_i\cdot p_j \nonumber
\end{eqnarray}
In the calculations it is convenient to take the  polarization vector
of the photon with momentum $p$ and helicity $\pm$ as follows
\begin{equation}
\epsilon^{\mu}_{\pm}(p,k)=\pm\, \frac{<p\pm|\gamma^{\mu}|k\mp>}
     {\sqrt{2}<k\mp|p\pm>}
\ee
with \be
\gamma_{\mu}  \epsilon^{\mu}_{\pm}(p,k)=\pm\, \frac{|p\mp><k\mp|+
|k\pm><p\pm|}{<k\mp|p\pm>} \ee
where $k$ is the reference momentum when judiciously chosen may simplify
the calculations significantly.
We find that choosing $p_2$ $(p_1)$ for photon helicity $+$ ($-$) as a
reference momentum (with the exception for the diagrams with $Z$ boson
exchanges for positive helicity of the electron, first two lines
below)  one obtains particularly simple analytical  expressions for the
helicity amplitudes. We find for the $Z$ diagrams:
\begin{eqnarray}
      A_Z^{++}&=&f_Z (g_v-g_a)\frac{<23>^2 <34>^*}{Z<25><15>} \nn \\
      A_Z^{+-}&=&f_Z (g_v-g_a) \frac{<14>^{*2} <34>}{Z<15>^*<25>^* } \nn \\
      A_Z^{-+}&=&-f_Z (g_v+g_a)\frac{ <13>^2 <34>^*}{Z<15><25>}\nn \\
      A_Z^{--}&=&-f_Z (g_v+g_a)\frac{<24>^{*2}<34>}{Z<15>^*<25>^*}
\end{eqnarray}
for $W$ diagrams
\begin{eqnarray}
      A_W^{-+}&=&-f_W \frac{<13>^2 <34>^*}{W_{14}<15><25>}\nn \\
      A_W^{--}&=&-f_W \frac{ <24>^{*2}<34>}{W_{23}<15>^*<25>^*}
\end{eqnarray}
and for $WW$ diagram
\begin{eqnarray}
      A_{WW}^{-+}&=&A_W^{-+}\frac{ <45>^* <23>^* <25>}{W_{23}<34>^*}\\
      A_{WW}^{--}&=&-A_W^{--}\frac{ <35> <14> <15>^*}{W_{14}<34>}
\end{eqnarray}
in the standard model, where the superscripts denote the helicity of the
electron and photon, respectively, $g_v=-1/2+2\sin^2\theta_W$, $g_a=-1/2$  and
\begin{eqnarray}
      f_Z&=&\frac{ieg^2}{\sqrt{2}\cos^2\theta_W} \nn \\
      f_W&=&-ieg^2\sqrt{2}\nn \\
      Z  &=&(p_3+p_4)^2-M_Z^2-iM_Z\Gamma_Z\nn \\
      W_{14}&=&(p_1-p_4)^2-M_W^2-iM_W\Gamma_W\nn \\
      W_{23}&=&(p_2-p_3)^2-M_W^2-iM_W\Gamma_W
\end{eqnarray}
For the additional contributions from $WW$ diagram with
anomalous $\delta\kappa$ and
$\lambda$ couplings we get
\begin{eqnarray}
      A_{\kappa}^{-+}&=&+\delta\kappa \,
            f_W\frac{<25>^* <45>^* <13>}{2W_{14}W_{23}}\\
      A_{\kappa}^{--}&=&-\delta\kappa \,
            f_W\frac{<24>^* <35> <15>}{2W_{14}W_{23}}\\
   A_{\lambda}^{-+}&=&\lambda\, f_W
       \frac{ <45>^*<14><24>^*<25>^*<23>}{2M_W^2W_{14}W_{23}}\\
   A_{\lambda}^{--}&=& -\lambda\, f_W\frac{<23>^*
            <35><13><14>^*<15>}{2M_W^2W_{14}W_{23}} \\
      \end{eqnarray}

\newpage
\noindent {\Large \bf Figure Captions}
\begin{itemize}
\item[Fig.1.] Cross section for the process $e^+e^-\rightarrow
\nu\bar{\nu}\gamma$ as a function of $\delta\kappa$ for $\lambda=0$
\item[Fig.2] Cross section for the process $e^+e^-\rightarrow
\nu\bar{\nu}\gamma$ as a function of $\lambda$ for $\delta\kappa=0$
\item[Fig.3.] Energy spectrum of the photon at $\sqrt{s}=500$ GeV for
the standard model and  for $\delta\kappa=-0.6$ and $2. $
\end{itemize}

\begin{thebibliography}{65}
\bibitem{conf} For a review and references see ``$e^+e^-$ Collisions at
500 GeV: The Physics Potential'', Proc.\ Munich, Annecy, Hamburg Workshop,
1991-1993, DESY Reports 92-123A+B, 93-123C
\bibitem{Karel} K.J.F. Gaemers, R.Gastmans,  F.M. Renard,
Phys.\ Rev.\ D19 (1979) 1605
\bibitem{frits} F.A. Berends {\em et. al.}, Nucl.\ Phys.\ B301 (1988) 583
\bibitem{HPZH} K. Hagiwara {\em at. al.}, Nucl.\ Phys.\ B282 (1987) 253
\bibitem{yehudai} E. Yehudai, SLAC-383, August 1991
\bibitem{CALKUL} F.A. Berends, W. Giele, Nucl.\ Phys.\ B294 (1987) 700
\bibitem{gk} J.F. Gunion, Z. Kunszt, Phys.\ Lett.\ B 161 (1985) 333
\bibitem{Hagizep} K, Hagiwara, D. Zeppenfeld, Nucl.\ Phys.\ B274 (1986) 1
\bibitem{He}X.G. He, B.H.J. McKellar, preprint UM-P-93/53, OZ-93/14
\bibitem{CLEO} E. Thorndike (CLEO Collab.), in Proc.\ of the American Physical
Society Meeting, Washington DC, April 1993
\end{thebibliography}
\end{document}